# Fluctuations of entropy and log-normal superstatistics

Sumiyoshi Abe[1,2,3]

[1]*Department of Physical Engineering, Mie University, Mie 514-8507, Japan*
[2]*Institut Supérieur des Matériaux et Mécaniques Avancés, 44 F. A. Bartholdi, 72000 Le Mans, France*
[3]*Inspire Institute Inc., Alexandria, Virginia 22303, USA*

**Abstract**   Nonequilibrium complex systems are often effectively described by the mixture of different dynamics on different time scales. Superstatistics, which is "statistics of statistics" with two largely separated time scales, offers a consistent theoretical framework for such a description. Here, a theory is developed for log-normal superstatistics based on the fluctuation theorem for entropy changes as well as the maximum entropy method. This gives novel physical insight into log-normal statistics, other than the traditional multiplicative random processes. A comment is made on a possible application of the theory to the fluctuating energy dissipation rate in turbulence.

PACS number(s): 05.40.-a, 05.20.-y, 05.70.-a

## I. INTRODUCTION

Complex systems in nonequilibrium states frequently exhibit hierarchical structures of dynamics that may be decomposed into different dynamics on different time scales. The decomposition is well defined if the time scales are largely separated from each other, and this is precisely the situation we are interested in, here. An illustrative example is a Brownian particle moving through a changing fluid environment with, e.g., temperature variations on a large spatial scale. There, two different dynamics are relevant. One is fast dynamics of the local Brownian motion of the particle, and the other is slow one describing the environmental variations. Consequently, the statistical property of the particle is given by the superposition of these two statistics, i.e., statistics of statistics, which is referred to as superstatistics.

Superstatistics has already been anticipated in Refs. [1-3]. To our knowledge, however, the work in Ref. [4] is the first, which explicitly considers the large separation of time scales as a physical basis.

Superstatistics typically yields stationary distributions of the non-Gaussian type [5]. Due to this fact, it has been examined for a variety of complex systems and phenomena including tracer particles in turbulent flows [6], velocity difference in turbulence [7], complex networks [8], ecosystems driven by hydro-climatic fluctuations [9], progression of cancer [10], and heterogeneous anomalous diffusion of viruses in cells [11].

Consider as an example a nonequilibrium system in an environment with slowly varying inverse temperature, $\beta$. This system can be thought of as the collection of many small spatial cells, each of which still consists of enough number of elements. After a certain duration of time, each cell may be in local equilibrium characterized by each value of $\beta$. Since there exist the changes of the local internal entropy, temperature gradient between two neighboring cells, and heat exchanges, consequently $\beta$ may vary in time. The time scale of this variation is naturally supposed to be much larger than that of relaxation to a local equilibrium state described by the following distribution:

$$p(\varepsilon_i | \beta) = \frac{1}{Z(\beta)} e^{-\beta \varepsilon_i}, \qquad (1)$$

where $\varepsilon_i$ is the *i*th value of the effective energy in the cell under consideration and $Z(\beta) = \sum_i e^{-\beta \varepsilon_i}$. Denoting the distribution of $\beta$ by $f(\beta)$, the superstatistical distribution to be observed on a long time scale is then given by

$$p(\varepsilon_i) = \int d\beta \, p(\varepsilon_i | \beta) f(\beta). \qquad (2)$$

In a special case when $\beta$ is global, i.e., spatially homogeneous but temporally varying, $p(\varepsilon_i | \beta)$ is regarded as the distribution describing the state of the whole system.

It is empirically known that superstatistics has three major classes [12]: namely, $\chi^2$, inverse $\chi^2$, and log-normal statistics for $f(\beta)$. Among others, log-normal superstatistics plays a pivotal role in the physics of turbulence [7].

In this paper, we develop a closed theoretical framework for log-normal superstatistics. Introducing the concept of entropy fluctuations and describing it by applying the fluctuation theorem [13-15] as well as the maximum entropy method, we show how the log-normal distributions are derived for a class of systems. We also make a comment on a possible application of the present approach to the problem of fluctuating energy dissipation rate in turbulence.

## II. FLUCTUATION THEOREM, MAXIMUM ENTROPY METHOD, AND LOG-NORAMAL SUPERSTATISTICS

First, let us discuss the formulation of superstatistics based on the conditional concepts in probability theory. In superstatistics, both the energy and inverse temperature are random variables, which are denoted by $E$ and $B$, respectively. $p(\varepsilon_i) \equiv p(E = \varepsilon_i)$ in Eq. (2) is a marginal distribution calculated from the joint distribution, $P(\varepsilon_i, \beta)$, which is, according to Bayes' rule, given by $p(\varepsilon_i | \beta) f(\beta)$, where $f(\beta)$ and $p(\varepsilon_i | B = \beta)$ are the marginal distribution of a slowly varying inverse temperature, $\beta$, and the conditional distribution of finding the system in the $i$th energy state, given the value of $\beta$, respectively.

The entropy associated with the joint distribution reads

$$S[E, B] = S[E | B] + S[B], \qquad (3)$$

where $S[E|B]$ and $S[B]$ are respectively given by

$$S[E|B] = \int d\beta \, f(\beta) \, S[E|\beta], \quad (4)$$

$$S[B] = -\int d\beta \, f(\beta) \ln f(\beta), \quad (5)$$

with

$$S[E|\beta] \equiv -\sum_i p(\varepsilon_i|\beta) \ln p(\varepsilon_i|\beta), \quad (6)$$

which is a function of $\beta$. In these equations, the Boltzmann constant is set equal to unity for the sake of simplicity. Note that $S[B]$ is determined up to an additive constant, since $f(\beta)$ has the dimension of $\beta^{-1}$ (as well-known for the entropy of a continuous distribution).

Eq. (4) explicitly shows the separation of the time scales: the summation over the fast variable, $E$, is taken first, and then the averaging over the slow variable, $B$, is performed afterward. That is, the randomness of $B$ is *quenched* in the calculation of $S[E|\beta]$. Substitution of Eq. (1) into Eq. (6) yields

$$S[E|\beta] = \beta U(\beta) + \ln Z(\beta), \quad (7)$$

where $U(\beta) = \sum_i \varepsilon_i p(\varepsilon_i|\beta)$ is the local internal energy. Therefore, we have the following expression for the conditional entropy:

$$S[E \mid B] = \overline{\beta \left[ U(\beta) - F(\beta) \right]}, \tag{8}$$

where $F(\beta) = -(1/\beta) \ln Z(\beta)$ is the local free energy and

$$\overline{Q(\beta)} \equiv \int d\beta \, f(\beta) \, Q(\beta). \tag{9}$$

Now, a primary question in superstatistics is how $f(\beta)$ can be obtained. $f(\beta)$ should be determined when a system is given, since physical properties of the system might become secondary if it would have to be derived from some additional probability-theoretic concepts such as the central limit theorem: it is desirable for $f(\beta)$ to be expressed in terms of physical quantities.

To answer this question, let us recall that slowly varying $\beta$ is due to the changes of the local internal entropy, temperature gradient, and heat exchanges. This fact allows the changes of the total entropy

$$\Delta S_{\text{tot}} = \Delta S[E \mid \beta] + \Delta S_s \tag{10}$$

in each cell to fluctuate, where $\Delta S[E \mid \beta] = S[E \mid \beta] - S_0$ and $\Delta S_s$ are the change of the local internal entropy from its fixed reference value, $S_0$, and the change of the entropy of the surroundings of the cell identified with the heat exchange, respectively. Then, the fluctuation theorem states that $\Delta S_{\text{tot}}$ obeys the following relation [15]:

$$\frac{P(-\Delta S_{tot})}{P(\Delta S_{tot})} = e^{-\Delta S_{tot}}, \tag{11}$$

where $P(\Delta S_{tot}) d(\Delta S_{tot})$ is the probability of finding the entropy change in the interval, $\Delta S_{tot}$ and $\Delta S_{tot} + d(\Delta S_{tot})$. Let us write $P(\Delta S_{tot})$ in the form

$$P(\Delta S_{tot}) = e^{\Omega(\Delta S_{tot})}. \tag{12}$$

Since $\Delta S_{tot}$ should not be so large, we expand $\Omega(\Delta S_{tot})$ up to the second order (i.e., the Gaussian approximation): $\Omega(\Delta S_{tot}) \cong a_0 + a_1 \Delta S_{tot} + (a_2/2)(\Delta S_{tot})^2$, where $a_n$ is the $n$th-order differential coefficient, $a_n = \Omega^{(n)}(0)$. The fluctuation theorem in Eq. (11) leads to

$$a_1 = \frac{1}{2}. \tag{13}$$

On the other hand, the normalizability of $P(\Delta S_{tot})$ requires $a_2$ to be negative:

$$a_2 = -2\lambda \qquad (\lambda > 0). \tag{14}$$

Accordingly, we have

$$\hat{f}(\beta) \equiv |J(\beta)| P(\Delta S_{tot}) \propto |J(\beta)| \exp\{\kappa' S[E|\beta] - \lambda S^2[E|\beta]\}, \tag{15}$$

where $J = dS[E|\beta]/d\beta$ is the Jacobian factor and $\kappa' = 2\lambda(S_0 - \Delta S_s) + 1/2$. $\hat{f}(\beta)$ in Eq. (15) is now identified with the distribution of $\beta$ in superstatistics.

In Eq. (15) and as we shall see below, *log-normal superstatistics is realized in the situation where the fluctuations of $\beta$ dominantly come from the fluctuations of the local internal entropy $S[E|\beta)$ and not from the heat exchanges.*

Before proceeding, we develop a further discussion about $\hat{f}(\beta)$ in Eq. (15). This distribution can also be obtained by the maximum-entropy-method approach to superstatistics. However, in marked contrast to the earlier works in Refs. [16-20], both the first and second moments of $S[E|\beta)$ should be constrained:

$$\overline{S[E|\beta)} = s, \qquad (16)$$

$$\overline{S^2[E|\beta)} = \sigma^2. \qquad (17)$$

Eq. (17) explicitly takes into account the fluctuations of entropy [21]. The maximum entropy method reads

$$\delta_f \left\{ S[B] - \alpha \left( \int d\beta\, f(\beta) - 1 \right) + \kappa'' \left( \int d\beta\, f(\beta)\, S[E|\beta) - s \right) \right.$$
$$\left. - \lambda \left( \int d\beta\, f(\beta)\, S^2[E|\beta) - \sigma^2 \right) \right\} = 0, \qquad (18)$$

where $S[B]$ is given in Eq. (5), $\{\alpha, \kappa'', \lambda\}$ the set of the Lagrange multipliers with $\alpha$ being the multiplier associated with the constraint on the normalization condition on $f(\beta)$, and $\delta_f$ stands for the variation with respect to $f(\beta)$. The stationary solution of this problem is given by $\check{f}(\beta) \propto \exp\{\kappa'' S[E|\beta) - \lambda S^2[E|\beta)\}$. This distribution is apparently different from $\hat{f}(\beta)$ in Eq. (15) by the Jacobian factor. However, as will be seen below, in a class of systems where log-normal superstatistics is realized, $S[E|\beta)$ behaves as $S[E|\beta) \sim \ln \beta$, leading to $|dS[E|\beta)/d\beta| \propto 1/\beta \sim \exp\{\kappa''' S[E|\beta)\}$, where $\kappa'''$ is a constant. Therefore, $\hat{f}(\beta)$ and $\check{f}(\beta)$ become identical after the redefinition of $\kappa'$.

Thus, we obtain the distribution of the following form:

$$\tilde{f}(\beta) = N \exp\left\{\kappa\, S[E|\beta) - \lambda\, S^2[E|\beta)\right\}, \tag{19}$$

where $\kappa$ and $\lambda$ are constants independent of $\beta$, and $N$ is a factor responsible to the normalization condition on $\tilde{f}(\beta)$.

Let us examine Eq. (19) for a simple superstatistical system consisting of $n$ mutually noninteracting Brownian particles with a common mass, $m$, which are put in a background fluid subject to large scale temperature variations. Given a local value of $\beta$, the conditional probability of finding the particle momenta $\{\mathbf{p}_1, \mathbf{p}_2, ..., \mathbf{p}_n\}$ in a cell is given by the multivariate Gaussian distribution:

$$p(\mathbf{p}_1, \mathbf{p}_2, ..., \mathbf{p}_n | \beta) = \frac{1}{Z(\beta)} \exp\left[-\beta\, \varepsilon(\mathbf{p}_1, \mathbf{p}_2, ..., \mathbf{p}_n)\right], \tag{20}$$

$$Z(\beta) = \frac{v^n}{n!} \left(\frac{2\pi m}{\beta h^2}\right)^{3n/2}, \tag{21}$$

where $\varepsilon(\mathbf{p}_1, \mathbf{p}_2, ..., \mathbf{p}_n) = (1/2m)(\mathbf{p}_1^2 + \mathbf{p}_2^2 + \cdots + \mathbf{p}_n^2)$, $h$ is the Planck constant, and $v$ the volume of the cell under consideration. The momenta are continuous random variables, and accordingly $S[E|\beta)$ in Eq. (6) should appropriately be modified, here. The result is known in the case of large $n$ as the Sackur-Tetrode formula [22]:

$$S[E|\beta) = n\left(-\frac{3}{2}\ln\beta + c\right), \tag{22}$$

where

$$c = \frac{3}{2}\ln\left(\frac{2\pi m}{h^2}\right) + \ln\left(\frac{v}{n}\right) + \frac{5}{2}. \tag{23}$$

Substitution of this formula into Eq. (19) gives rise to

$$\tilde{f}(\beta) = \frac{3n}{2\beta}\sqrt{\frac{\lambda}{\pi}}\exp\left[-\frac{9\lambda n^2}{4}(\ln\beta - \ln\beta^*)^2\right] \tag{24}$$

with

$$\ln\beta^* = \frac{2c}{3} + \frac{2}{9\lambda n^2} - \frac{\kappa}{3\lambda n}. \tag{25}$$

Superposition of the distribution in Eq. (20) with respect to $\tilde{f}(\beta)$ in Eq. (24) yields log-normal superstatistics for a non-Gaussian distribution, the explicit form of which is not mathematically available.

Thus, we have formulated log-normal superstatistics based on both the fluctuation theorem for entropy changes and the maximum entropy method in the self-consistent manners.

We wish to emphasize that the present approach is radically different from the traditional multiplicative random processes for log-normal statistics. A crucial point here is that *B* does not have to be a geometric mean of independently and identically distributed (i.i.d.) positive random variables. Essentially, the absence of correlations is not premised, here.

### III. COMMENT ON FLUCTUATING ENERGY DISSIPATION RATE IN TURBULENCE

The phenomenon of turbulence is an important physical example, in which non-Gaussian distributions are observed. In this section, we examine the discussion developed in Sec. III for fluctuating energy dissipation rate and velocity difference statistics in turbulence.

A classical picture of fully developed turbulence contains cascading from large eddies to smaller ones. A key quantity in this process is the fluctuating energy dissipation rate, $\varepsilon_r$, defined in a cell with spatial scale *r*. [This notation for the energy dissipation rate should not be mixed with that for the energy in Eq. (1).] A basic assumption is that cascading is a multiplicative process and each $\varepsilon_r$ is an i.i.d. random variable. That is, the distribution of $\varepsilon_r$ is taken to be of the log-normal form [23,24].

This classical picture is still practically supported well by both experiments and numerical simulations [25-27] (although it may violate the inequality to be satisfied by the exponent of the structure function [28], in general).

A stationary turbulent flow is characterized by the distribution of longitudinal velocity difference over distance $r$: $\delta v_r = \mathbf{e}_r \cdot [\mathbf{v}(\mathbf{x}+\mathbf{r}) - \mathbf{v}(\mathbf{x})]$, where $\mathbf{v}$ and $\mathbf{e}_r$ are the velocity field and the unit vector in the direction of separation, respectively. Let $p(\delta v_r | \varepsilon_r)$ be the conditional velocity difference distribution, given a value of $\varepsilon_r$. It is discussed in Refs. [7,29] that in the symmetric case (i.e., the absence of skewness) the Gaussian form

$$p(\delta v_r | \varepsilon_r) = \frac{1}{\sqrt{2\pi(r\varepsilon_r)^{2/3}}} \exp\left[-\frac{(\delta v_r)^2}{2(r\varepsilon_r)^{2/3}}\right] \qquad (26)$$

is in accordance with experimental observations. Therefore, denoting the distribution of $\varepsilon_r$ by $f(\varepsilon_r)$, we have for the velocity difference distribution the following superstatistical expression:

$$p(\delta v_r) = \int d\varepsilon_r \, f(\varepsilon_r) \, p(\delta v_r | \varepsilon_r). \qquad (27)$$

Now, let us consider $f(\varepsilon_r)$ within the framework developed in Sec. II. Here, $\delta V_r$ (the value of which is $\delta v_r$) and $\varepsilon_r$ correspond to $E$ and $\beta$, respectively. Thus, we have

$$S[\delta V_r | \varepsilon_r] = -\int d(\delta v_r)\, p(\delta v_r | \varepsilon_r) \ln[p(\delta v_r | \varepsilon_r) v_0]$$

$$= \frac{1}{3}\left(\ln \varepsilon_r + c_0\right), \tag{28}$$

where $v_0$ is a constant having the dimension of $\delta v_r$ and

$$c_0 = \ln \frac{r}{v_0^3} + \frac{3}{2}[1 + \ln(2\pi)], \tag{29}$$

provided that the dimensionality of the first term on the right-hand side of Eq. (29) should be understood in conformity with Eq. (28). Therefore, the distribution corresponding to Eq. (19) is calculated to be

$$\tilde{f}(\varepsilon_r) = \frac{1}{3\varepsilon_r}\sqrt{\frac{\lambda}{\pi}} \exp\left[-\frac{\lambda}{9}(\ln \varepsilon_r - \ln \varepsilon^*)^2\right] \tag{30}$$

with

$$\ln \varepsilon^* = \frac{3\kappa}{2\lambda} + \frac{9}{2\lambda} - c_0. \tag{31}$$

Substitution of this distribution into $f(\varepsilon_r)$ in Eq. (27) yields log-normal superstatistics for turbulence, which is consistent with empirical data in the symmetric case [7,30]. In the present picture, the fluctuations of the energy dissipation rate are essentially due to fluctuations of the local entropy.

Finally, we wish to point out that the above discussion leads to an interesting question if the Gaussian fluctuations of the entropy change can experimentally be observed in the phenomenon of turbulence.

## IV. CONCLUSION

We have developed the theory of log-normal superstatistics based on the concept of entropy fluctuations described by both the fluctuation theorem and the maximum entropy method. We have presented a novel approach to the log-normal distribution, which has traditionally been discussed for multiplicative random processes. We have applied this theory to the problem of stationary symmetric turbulence.

In the present work, we have treated only analytically-tractable systems. To further examine the idea, it is of interest to consider, for example, skewness in turbulence. Such a problem will require deviation from exact log-normal statistics.

## ACKNOWLEDGMENT

This work was supported in part by a Grant-in-Aid for Scientific Research from the Japan Society for the Promotion of Science.